# Open Access beyond cable: The case of Interactive TV

Hernan Galperin
University of Southern California

François Bar
Stanford University

**- draft (please do not cite or quote) -**

Abstract

In this paper we analyze the development of interactive TV in the U.S. and Western Europe. We argue that despite the nascent character of the market there are important regulatory issues at stake, as exemplified by the AOL/TW merger and the British Interactive Broadcasting case. Absent rules that provide for non-discriminatory access to network components (including terminal equipment specifications), dominant platform operators are likely to leverage ownership of delivery infrastructure into market power over interactive TV services. While integration between platform operator, service provider and terminal vendor may facilitate the introduction of services in the short-term, the lasting result will be a collection of fragmented "walled gardens" offering limited content and applications. Would interactive TV develop under such model, the exciting opportunities for broad-based innovation and extended access to multiple information, entertainment and educational services opened by the new generation of broadcasting technologies will be foregone.

The broadcasting industry is rapidly entering the era of digitization, distributed intelligence, and interactivity. Despite lingering standards issues, digital transmission is replacing analog in the three major delivery platforms (terrestrial, cable, and DBS). Powerful user terminal built upon PC hardware and software technology are replacing dumb analog TV sets. And after several failed attempts in the past, interactive TV services are finally poised for large-scale deployment. This transition opens many exciting opportunities for new applications and services, ranging from TV-based electronic commerce (known as "t-commerce") to interactive educational programming. These will evolve as broadcasters, software vendors, hardware makers, and content creators and users experiment with novel ways to enhance and perhaps transform the TV experience entirely.

This transition, however, also raises several questions about who will shape the architecture of the emerging broadcasting networks, and hence determine business models, communication patterns, and the dynamics of technological innovation for the next generation of broadcasting. Will programmers or network operators decide which interactive services will be made available to users? Will electronic marketplaces develop as open transactional spaces or "walled gardens"? Will network users be able to connect new terminal equipment to the network and experiment with new network uses (e.g., peer-to-peer applications)? In lieu to these questions policymakers are faced with several pressing concerns. How to create incentives for firms to invest in this infant marketplace and at the same time protect competition in services and applications, foster decentralized innovation, and guarantee users' access to a wide range of information and transaction services? Would ex-ante regulation squelch the success of a sector that, after many failed attempts, is now ready for prime-time? What regulatory principles and tools should be used to confront the questions raised by interactive TV?

Far from hypothetical, these questions have already surfaced in several high-profile cases, in particular the AOL/Time Warner merger, which combined the world's largest ISP and early entrant in the interactive TV market with the U.S.' second largest cable operator and major worldwide programmer. In reviewing the merger, the Federal Trade

Commission (FTC) and the Federal Communications Commission (FCC) found that the combination of distribution facilities, service operations, and content held by AOL/Time Warner raised competition concerns in three markets: broadband Internet access service, broadband Internet transport service, and interactive TV. While regulators imposed several merger conditions concerning broadband Internet access and transport services, those relating to interactive TV were, in comparison, rather minor.

In this paper we analyze the development of interactive TV in the U.S. and Western Europe and the policy debates that have accompanied it. We argue that despite the nascent character of the market there are important regulatory issues at stake that will determine the future architecture of this information distribution platform. Absent rules that provide for non-discriminatory access to network components and a degree of standardization for terminal equipment, dominant platform operators are likely to leverage ownership of delivery infrastructure into market power over interactive TV services. While in the short-term integration between platform operator, service provider and terminal vendor is likely to facilitate the introduction of services, the lasting result will be a collection of fragmented "walled gardens" offering limited content and applications. Would interactive TV develop under such model, the exciting opportunities for broad-based innovation and extended access to multiple information, entertainment and educational services may be foregone.

We recognize that given the incipient nature of the market (particularly in the U.S.), it would be premature for regulators to attempt to implement detailed industry-wide rules for interactive TV platforms and services. There is simply too much uncertainty about which services users will want (and at what price), how the technology will evolve, and what business models will emerge. It may be that the dynamics of market competition will stimulate a migration from proprietary technologies and "walled garden" business models to open standards and interconnected networks (much like what happened in the narrow-band ISP industry during the 1990's), thus making regulatory safeguards less necessary. We nonetheless content that it is not too early to establish general rules and first principles with which to monitor market developments. Interactive TV provides another test case of how to adapt existing broadcasting and telecommunications regulation to the ongoing convergence of digital networks in a way that balances industry development with the economic and social benefits associated with open network access. The debate over broadband cable Internet offered a first approach to the problem and some important lessons.[1] While technologies may vary from case to case, our ultimate policy goals should not.

The case of interactive TV thus offers an opportunity to investigate how desirable policy goals, among them competition, broad-based innovation, and widespread access to information "from diverse and antagonistic sources",[2] should be implemented in the post-convergence environment. We start the paper by reviewing the evolution of the broadcasting industry through three successive models: the "Fordist" TV model, the multichannel TV model, and the emerging interactive TV model. Second, we characterize the basic components of interactive TV and explore the concerns raised by the evolution of multichannel video programming distributors (MVPDs) into interactive TV platform operators. Our conclusion is that dominant MVPDs are likely to have the ability and the incentive to leverage control over the transmission platform into the ITV applications environment, engineering market outcomes in favor of affiliated programmers, electronic retailers, and ITV service providers. We note that unlike the case of broadband cable Internet the concerns go beyond infrastructure access control and include the use of proprietary terminal equipment standards.

Third, we review how regulators in the U.S. and the European Union (EU) have so far responded to these concerns by contrasting two cases: the AOL/Time Warner merger and British Interactive Broadcasting case. We conclude that the wait-and-see approach taken by American regulators risks tolerating the deployment of a closed architecture network that would inhibit healthy competition in interactive TV services, hamper innovation, and create a large group of second-class digital economy citizens with access to a very limited array of entertainment, transaction, and educational services. We also note that the imposition of limited open access requirements in the UK market has hardly hampered investments in interactive TV. Finally, we outline a general framework for regulatory thinking about open network access that reflects the convergence of communications industry sectors and the need to balance seemingly conflicting policy goals.

---

[1] See Bar et al. (2000).
[2] Associated Press v. Unites States, 326 U.S. 1, p. 20 (1945).

## 1. The three generations of broadcasting

The half-century old broadcasting industry has developed through three technological generations, each characterized by different types of services, business models, control strategies, and regulatory environment (Table 1). It is interesting to note that each new generation has not thoroughly overhauled the pre-existing industry structure, but rather added a layer of complexity to it. From the start of commercial broadcasting in the post-war period to about the mid-1970's, television consisted essentially of one-way terrestrial broadcasting of a limited number of channels which aggregated and sold large audiences to advertisers, the operators of which were protected by rules that restricted competition both within the industry and from new entrants (Horwitz, 1989; Hazlett, 1998). The regulatory model was based on the idea that broadcasters (both public and private) are trustees of a public resource (the radio spectrum) and thus under obligation to serve the public interest as defined by the government. While government protection from competition ensured the profitability of most broadcasting operations, fulfillment of public interest obligations was, at best, questionable.

During the 1970's, a series of technological and regulatory developments created the conditions for the rapid growth of cable, and later direct-to-home satellite TV (DBS).[3] These new platforms essentially offered more (and today, much more) of the same service: one-way delivery of branded packages of TV programming. However, a new business model emerged based on the collection of payments directly from subscribers, spawning the growth of specialized channels with limited audience base. The regulatory model was fashioned as a mix of traditional broadcasting and utility regulation. Cable operators were for the most part granted monopoly franchises by local authorities in return for payments and limited access obligations (the so-called PEG and leased access channels). The federal government later imposed restrictions on cable operators' editorial control by limiting the number of channels that can be occupied by affiliated video programmers.[4] Notwithstanding these restrictions, cable essentially developed as a closed network with tight integration between network layers (transmission infrastructure, service provision, and terminal equipment).

---

[3] For a discussion of these developments see Owen (1999).

The regulatory model for the second generation of broadcasting thus evolved in remarkable contrast with that of the telecommunications network, particularly after the FCC reversed its defense of "network integrity" in the late 1960's, encouraging open attachment of terminal devices,[5] network interconnection,[6] and third-party access to unbundled network elements.[7] It is thus no surprise that while the telecommunications industry has experienced a period of unprecedented innovation based on experimentation by network users and third-party service providers over the last three decades, the pace of innovation and introduction of new services in the cable industry has, in comparison, been limited by the creativity and the narrow economic incentives of those in quasi-monopoly control of the transmission infrastructure.

After much delay, the revolution in digital processing and transmission of information is finally ushering the broadcasting industry into a new era. As a long-time industry analyst described it, "after a half-century of glacial creep, television technology begun to change at the same dizzying pace as the wares of Silicon Valley."[8] But as MVPDs evolve from distributors of video programming into operators of a network that supports a variety of information services, regulators are confronted with a fundamental policy question: under what regulatory model should the next generation of television services develop? That of the second generation of broadcasting - even though cable and satellite operators may effectively act as providers of telecommunications infrastructure rather than as content aggregators and distributors - or that of open network access that has guided much telecommunications policy over the last decades?

---

[4] See 47 U.S.C. § 613(f).
[5] See In the Matter of Use Of The Carterfone Device In Message Toll Telephone Service, Docket No. 16942; 13 FCC 2d 420 (1968).
[6] See Specialized Common Carrier Services, 29 FCC 2d 870 (1971).
[7] See In the Matter of Regulatory Policies Concerning Resale and Shared Use of Common Carrier Services and Facilities, 60 FCC 2d 261 (1976).
[8] Owen (1999), p.3.

**Table 1: The three generation of broadcasting**

|  | 1st Generation: Fordist TV | 2nd Generation: Multichannel TV | 3rd Generation: Interactive TV |
|---|---|---|---|
| Service | One-way broadcasting of few video channels | One-way broadcasting of multiple video channels | Two-way delivery of multiple video channels and other services |
| Business model | Mass advertising and/or license fees | Mass advertising, license fees, and subscriptions | Targeted advertising, subscriptions, and transaction fees |
| Control strategies | Property rights over spectrum license | Integration of distribution and content assets | Access control and proprietary standards |
| Regulatory model | Public trustee (incumbent protection) | Mix of public trustee and limited utility regulation | (to be defined) |

## 2. What is interactive TV?

Due to the infancy of the market, any description of what constitutes interactive TV is necessarily a working definition. In the context of this paper, we adopt a definition broader in scope than the one recently proposed by the FCC. In a public consultation, the Commission has defined interactive TV as "a service that supports subscriber-initiated choices or actions that are related to one or more video programming streams."[9] Instead, we agree with the British broadcasting regulator (the Independent Television Commission, henceforth ITC) in that interactive TV services are "pull" services initiated by the subscriber to a MVPD that are *not* necessarily related to any specific video programming.[10] In fact, we differentiate between two essentially different types of interactive TV services: dedicated and program-related services.

---

[9] FCC (2001). In the Matter of Nondiscrimination in the Distribution of Interactive Television Service over Cable. CS Docket No. 01-7, p. 2.
[10] See Independent Television Commission (2000). Interactive Television: An ITC Public Consultation. Available at www.itc.org.uk.

The first are stand-alone services not related to any specific programming stream. Typically, this will be entertainment, information, and transaction services provided by electronic retailers on the basis of contracts with the MVPD, which acts essentially as platform operator, offering subscribers a "window" for t-commerce. Examples of these services already available or in the deployment stage are Electronic Programming Guides or EPGs (in a sense the more mature of interactive TV services, discussed in more detail below), video-on-demand (already launched in several cable markets by interactive TV provider DIVA), e-mail (offered for example by Microsoft's WebTV), games, gambling, and electronic banking (available for example through the BiB platform). While some of these electronic retailers may already have Internet-based services, these typically need to be re-authored for the different systems used by network operators (though as discussed below there are several standardization efforts under way). In comparison, these platforms are strictly "walled gardens" environments: the network operator selects a limited number of electronic merchants that is made available to users, and typically charges an upfront fee for access control (e.g., authentication) and billing services as well as a commission on sales.[11]

Program-related services refer to interactive TV services that are directly related to one or more video programming streams. These services allow users to obtain additional data related to the content (either programming or advertising), to select from a menu of video feeds, to play or bet along with a show or sports event, to interact with other viewers of the same program, or to initiate transactions of goods or services featured in the video programming. In this case, interactive TV enhancements (such as ATVEF "triggers")[12] are overlayed onto the MPEG video programming stream. These enhancements, when selected, direct viewers to content stored either in the set-top box or a remote server. In the latter case, the enhanced content is delivered either through the same video pipeline or a separate transmission line (e.g., an Internet connection). Examples of services already available or in the deployment stage are the delivery of on-demand financial information and stock quotes along with a business new channel (e.g., CNBC in partnership with Wink Communications), enhanced TV commercials that allow viewers to request more information about the product, enhanced educational programming (a partnership between PBS and RespondTV), and services that allow users to play or bet along quiz shows, reality shows, and live sports events.

---

[11] For example, BiB reportedly takes a 8% commission on sales. See "At least television works," The Economist, October 7, 2000.

[12] ATVEF (Advanced Television Enhancement Forum) is a cross-industry group formed by programmers, broadcasters, interactive TV service providers, hardware makers, and software developers intended to create standard protocols for the delivery of interactive TV enhancements. See www.atvef.com.

In the case of program-related services the programmer or advertiser will typically contract with an interactive TV service provider for the creation of programming enhancements, storage of interactive content, and management of return channel data. Nevertheless, the compliance of the network operator is still needed to deliver the downstream program enhancements, to allow compatibility between interactive TV applications and operator-provided customer equipment (unless a stand-alone box is used, which is unlikely for reasons discussed below), and possibly to provide the high-speed return path needed for certain applications. Therefore, the ability of programmers and interactive TV service providers to experiment with and deploy services is *de facto* dependent on access to both the transmission infrastructure and the home terminal functions. As we argue in the next section, unless regulatory safeguards guarantee such access on non-discriminatory terms, the next generation of broadcasting services will (like the previous generations) be characterized by slow innovation and limited competition.

While the market for interactive TV is still maturing, the pace of development has accelerated dramatically in recent years. Growth has been fueled by decreasing equipment costs (of both network hardware and home terminals)[13] and related infrastructure investments that facilitate the provision of interactive TV services, in particular the slow but steady migration to digital transmission standards in terrestrial, cable, and satellite TV.[14] Analysts estimate that in Britain 25% of households will use some interactive TV service this year, with the U.S. lagging behind with 7.5%.[15] Furthermore, it is expected that TV-based Internet access will eventually outpace PC-based access, particularly in regions with low PC penetration such as Europe.[16] Such rapid growth should not preclude regulatory action as some have argued (e.g., Elhauge, 2001). Quite the contrary, it represents an opportunity for policymakers to shape the next generation of broadcasting services in a way that fosters innovation and competition during the formative period of the industry, much like policy intervention favoring open network access in telecommunications networks starting in the late 1960's made possible the Internet revolution several years later.[17]

---

[13] For example, the cost of video servers, the core component of video-on-demand systems, has dropped 90% over the last years (see "Interactive Television: Fulfilling the Promise," Broadcasting & Cable, July 10, 2000, p.22). The cost of digital set-top boxes has also dropped dramatically in recent years following the decline in prices for computer components.

[14] For a discussion of the transition to digital TV see Galperin (forthcoming).

### 3. Policy concerns raised by interactive TV

Different opportunities for dominant network operators to foreclose competition in the adjacent markets for interactive TV and video programming services exist at different network layers. In this section we examine the opportunities and incentives for anti-competitive behavior that result from vertical integration between network operators and interactive TV service provider at three network layers: the transmission system, the return path, and the home terminal (typically a digital set-top box). We argue that switching costs, network complementarities, first-mover advantages and technical advantages are likely to create a dominant platform for the delivery of ITV services in every geographic market (in most cases the monopoly cable franchisee) for which there will be no close substitutes. As a result, unaffiliated interactive TV service providers and programmers may face discriminatory access to the transmission infrastructure necessary to compete in the third generation of broadcasting. This risk will likely dampen innovation and discourage entry by third-party application developers and programmers into the interactive TV market.

---

[15] Source: Jupiter Media Matrix.
[16] The British government has already stated its goal that following the digital switchover every home with a TV-set and a telephone line has access to the Internet. See Joint ITC, OFTEL, and OFT Advice to Government on Digital Television, November 2000, available at www.oftel.gov.uk.
[17] See Bar et al. (2000).

a) Transmission system

In the case of program-related services, the most apparent opportunity that exists for network operators to discriminate in favor of affiliated interactive TV service providers and programmers consists in "stripping" the interactive TV enhancements from the video signal of an unaffiliated programmer, thus blocking access to the enhanced features offered by the competing programmer. Time Warner Cable, for example, has repeatedly blocked subscriber access to Guide Plus+, a free EPG offered by interactive TV provider Gemstar that is carried over the VBI.[18] By stripping out the data inserted by Gemstar in the VBI of local television broadcast stations, Time Warner was favoring a competing EPG offered by its cable subsidiaries.[19]

It is important to note that in the case of program-related services, the issue is not of programmers' access rights to cable distribution per se. Even when the network operator has agreed (or is forced by statute, as in the case of local TV stations) to carry an unaffiliated programmer, it has the ability to favor its own related programmer (e.g., AOL/TW's Cartoon Network vs. Disney's Disney Channel) by stripping out the interactive features of a rival's video signal (e.g., the ATVEF triggers). Alternatively, the platform operator can slow down the rate of transmission of the downstream interactive data, thus interfering with the synchronization between the interactive service and the programming to which it is related. The ultimate effect is similar: to make an unaffiliated video signal less compelling as an information/entertainment experience.

In the case of dedicated interactive TV services, the bundling of transmission and interactive TV services presents question similar to those discussed in the context of the debate over broadband cable Internet. Nonetheless, the concerns are exacerbated by the fact that, unlike ISPs, interactive TV service providers face from the start the closed network architecture of the second generation of broadcasting, rather than the open, end-to-end architecture of the first-generation Internet. If a single transmission platform emerges as the only viable alternatively to compete in the provision of interactive TV services (an assumption we explore below), the platform operator does not need to reengineer the network in order to favor its affiliates because entry will be, from the outset, by invitation only. As the ITC explains,

> "The distinctiveness of interactive television services as compared with the Internet is manifested in the "walled garden" concept, where a limited number of sites or parts of sites are selected by the interactive licensee (…). In this environment an interactive licensee has the potential to exercise a degree of pre-selection and control of content through their contractual relationships with the providers of the walled garden content. This factor (…) suggest that a somewhat different treatment is needed than applied to the Internet" (ITC, 2000, p. 7).

---

[18] The Vertical Blanking Interval (VBI) is the interval between television frames in analog broadcasting, which allows for a limited data transmission capacity (typically used for closed-captioning services).
[19] See In Re Petition for Special Relief of Gemstar, CSR-5528-Z. Due to regulatory scrutiny of the AOL/Time Warner merger, Time Warner Cable has reportedly ceased such practice.

Critics of ex-ante regulatory action on interactive TV nonetheless contend that network operators are unlikely to have incentives to discriminate against unaffiliated interactive TV providers or programmers, and thus any rules imposed will have costly effects on investments and service efficiency. As Elhauge argues,

> "it is implausible that any local ITV platform could hope to raise entry barriers by denying access to rival ITV service providers. Because it could not raise entry barriers, it would have incentives to deny access if and only if such a denial were efficient: either because the denied provider would not efficiently fit the platform or because vertical integration of ITV platforms and services is more efficient. Any interference with such decisions would make ITV markets inefficient, with higher costs or lower quality for consumers" (Elhauge, 2001, p. 35).

In our opinion, the argument that vertically-integrated network operators will lack incentives to discriminate, and hence will offer access to as many programmers and service providers as would "fit the platform," is unconvincing. While it is clear that a network operator will want to maximize available content in order to attract subscribers, it is not clear that it will have incentives to grant users access to competing interactive TV service providers, particularly given the existence of close substitutes in the content market. If the revenue gains from additional nationwide sales from affiliated programmers and interactive TV service providers (in the form of carriage fees, advertising, commissions on t-commerce, etc.) exceeds the foregone revenue from lost subscribers within the cable's franchises (i.e., those who demanded the content or services not available), discrimination maximizes profits for the vertically-integrated operator.[20] Discrimination may also help preserve cable's advantage over competing MVPDs in the long-term as long as it discourages entry by third parties in the upstream market, thus making alternative MVPDs more dependent on programming and interactive TV services supplied by cable's affiliates.

The argument that vertically-integrated network operators will lack incentives to discriminate is based on a static notion of market efficiency and consumer welfare that overlooks two fundamental goals in communication policymaking: that of fostering dynamic innovation in broadcasting services and that of promoting widespread access to information "from diverse and antagonistic sources." The very existence of a gatekeeper between interactive TV services and end-users will suffice to discourage entry by application developers and programmers.[21] The ultimate result would be an efficient (in static terms) but highly constrained environment for the conduct of commerce and speech. In addition, the argument simply contradicts the actual evidence of discriminatory behavior by cable operators against unaffiliated interactive TV service providers. As discussed below, the AOL/TW merger case offered ample evidence of the use of control over the cable transmission infrastructure to squelch competition in the market for next-generation broadcasting services.

b) Return path

Interactive TV is based on the presence of a return path that provides upstream communication between the home terminal and the service provider. This return path can potentially take many forms: it may be a standard dial-up Internet connection (used for example by WebTV), a proprietary version of a dial-up connection (used by AOLTV), an "out-of-band" reverse data channel (as used by most cable operators), or even a wireless two-way radio connection (used by Gemstar's GuidePlus+). For most dedicated interactive TV services, the speed and synchronization of the return path with the video signal do not pose significant market entry barriers. However, for program-related services and dedicated services that do not tolerate latency or require full screen video streaming, the availability of a high-speed, high capacity return path that works in close coordination with the related video feed is be essential to create a compelling interactive TV experience. In most cases this is likely to be a broadband Internet connection.

---

[20] See Declaration of Gregory Sidak and Hal Singer In the Matter of Nondiscrimination in the Distribution of Interactive Television Services over Cable. CS Docket No. 01-7. The authors demonstrate that this is the case when discrimination discourages unaffiliated programmers or content providers from entering the market or induces exit by existing ones.
[21] See Lemley and Lessig (2000).

As a result, cable is likely to become the dominant platform for interactive TV services, at least in the U.S. where the cable plant is already installed and rapidly being upgraded to provide two-way digital services (we discuss below the European case where DBS seems to have a first-mover advantage over cable). As the FTC explains, "cable has distinct advantages over alternative ITV transport methods. The television signal is already transmitted over cable, which makes synchronizing viewer interaction with the programming easier. Neither satellite nor DSL connections can integrate the cable video programming and the interactive functionality as smoothly as cable."[22] Cable networks also provide extensive transmission capacity in both directions (downstream and upstream), a critical factor for the new generation of broadcasting services. Furthermore, operators have already made substantial investments in upgrading facilities to offer digital TV packages and broadband cable Internet, upon which interactive TV services could be piggy-backed.[23] As the FCC concludes, "our understanding of the current state of technology suggests that the cable platform is likely to be the best suited for delivering ITV services, particularly high speed services, for at least the near term."[24]

---

[22] Federal Trade Commission (2000). Complaint In the matter of America Online Inc. and Time Warner Inc.. Docket No. C-3989, p. 4.
[23] The NCTA (National Cable & Telecommunications Association) estimates that by the end of 2001 60% of households will be passed by two-way cable plants.
[24] Federal Communication Commission (2001). In the Matter of Nondiscrimination in the Distribution of Interactive Television Service over Cable. CS Docket No. 01-7, p. 8.

The lack of a credible competitor to discipline cable operators opens several avenues for discriminatory behavior in favor of affiliated programmers and interactive TV service providers. Cable operators can simply refuse to provide a return path to third parties. In fact, during the AOL/TW merger review the FCC received several complaints about Time-Warner Cable's refusal to provide guarantees about non-discriminatory use of the return path from unaffiliated programmers.[25] The network operator may also degrade the quality of the return path (in terms of speed or reliability) offered to third parties. In addition, it could seek charges for t-commerce transactions originated through its platform. This would be similar to an ISP seeking compensation from electronic retailers such as Amazon.com for every item sold to its subscribers. Rather than simply enabling transactions under the end-to-end principle, the transport operator would erect a tollgate between buyers and sellers.[26] Lastly, valuable customer data can be obtained from the return path even when the platform operator is not a party of the commercial transaction taking place. This has raised concerns not only from third-party programmers and interactive TV service providers but also from consumer groups concerned about viewers having little control over how the return-path data will be compiled and used.[27]

c) Home terminal

The third necessary component of an interactive TV system is the home terminal or digital set-top box. As the number and complexity of interactive TV services increases, so will the processing and storage capacity of the home terminal in order to perform the different tasks. In essence, a digital set-top box is similar to a stripped-down PC. There are at least two software components within the digital set-top box that, absent regulatory safeguards or open industry standards, present opportunities for discriminatory behavior by dominant platform operators. The first is the Application Program Interface (API), which is the software layer between the operating system (e.g., Microsoft's Microsoft TV or Liberate's TV Navigator) and the different applications running on the terminal. In order to be available to users, interactive TV applications need to interact with the set-top API, in the same way a word processor interacts with the PC's operating system.

---

[25] See Ex Parte submission of The Walt Disney Company to the FCC filed October 25, 2000, in CS Docket No. 00-30.
[26] As a Time Warner Cable executive explained, "if a programmer wants to offer its advertisers the ability to have two-way communications with viewers, the cable operator has to be part of that." See "AOL-Time Warner rivals preparing for interactive TV fight." The New York Times, September 11, 2000, p. C1.
[27] See TV watches you: The prying eyes of interactive TV. Center for Digital Democracy, June, 2001. Available at www.democraticmedia.org.

Unlike the more mature PC industry, there is no de facto industry standard for set-top box APIs. If such a standard were to develop in the future, and if its technical specifications were available to application developers on non-discriminatory terms, the competitive concerns associated with the API would be mitigated. There are a number of industry consortia working to create an open platform for interactive TV. Among them are OpenCable's OpenCable Applications Platform (OCAP), the DVB's Multimedia Home Platform (MHP), and even a Linux-based platform sponsored by the TV Linux Alliance.[28] For the foreseeable future, however, proprietary (i.e., non-interoperable) APIs will be the deployed by network operators, forcing developers to rewrite interactive TV applications for several different environments.

In order to enter the market an interactive TV service provider (assuming it has secured both downstream and upstream carriage) faces two options: it can either contract with the dominant platform operator to gain access to the installed base of terminals, or it can deploy a stand-alone box, thus bypassing the proprietary terminal components altogether. The latter option, while theoretically possible, is nonetheless uneconomical for most potential entrants. It is highly unlikely that users will be willing to buy a new box for every new interactive TV application (who would be willing to buy a separate PC for every new application?). The failure of stand-alone boxes marketed by companies like ReplayTV (which allowed digital video recording) and WebTV (despite heavy marketing spending by its parent Microsoft) has shown that consumers prefer a single box that integrates traditional video programming with new services.[29] Second, the evidence from the introduction of DBS, wireless telephony, and digital TV shows that heavy terminal subsidies are necessary. Thus, as an European competition official explains, "the scale of investment required means that the new entrants' most realistic option is to provide a service using the set-top boxes which already exist."[30]

---

[28] OpenCable is an initiative of CableLabs, a Research and Development consortium formed by U.S. cable operators. The DVB (Digital Video Broadcasting) group is an European consortium formed by equipment manufacturers, broadcasters, content producers, software developers, and representatives of national regulatory bodies. The TV Linux Alliance is a U.S.-based consortium of technology suppliers to cable, satellite and telecommunications network operators.

[29] In fact, Personal Video Recorders (PVRs) by ReplayTV and Tivo are now being embedded into cable and satellite receivers. See "ReplayTV goes further upscale", Red Herring, September 5, 2001.

Access to the API specifications and related facilities (authoring tools, authorization keys, memory control, etc.) is therefore critical for potential entrants in the interactive TV services market. This creates several opportunities for strategic behavior by dominant network operators such as refusing to provide authoring tools, discriminatory access pricing, discriminatory allocation of set-top boxes facilities (e.g., set-top box memory for caching), and bundling of API access with other services (e.g., conditional access or subscription management). European competition authorities have even framed the issue as a problem of access to an essential telecommunications facility, which under European law triggers several non-discriminatory obligations for dominant network operators.[31] While this doctrine is yet to be applied in a case regarding interactive TV, the language of the new regulatory framework proposed by the European Commission (EC) reveals the intent to extend existing access obligations to the set-top box components controlled by dominant cable and satellite TV operators.[32]

The second component that raises policy concerns is the Electronic Programming Guide (EPG), a navigation software that allows users to browse and select TV channels and services. With the manifold increase in the number of channels and applications made possible by the transition to digital TV, the EPG is expected to become to the broadcasting industry what Web portals have become to the Internet: powerful tools to direct traffic and obtain advertising revenues (Mansell, 1999). From a regulatory standpoint, the main concern is that dominant platform operators do not use the EPG to leverage their power onto the market for video programming and interactive TV services. As European regulators explain,

---

[30] McCallum (1999), p. 11.
[31] See Temple Lang (1997).
[32] See Amended proposal for a Directive of the European Parliament and of the Council on access to, and interconnection of, electronic communications networks and associated facilities. COM(2001) 369 final.

> "Issues of ensuring listing of third-party services or programming, and the quality of such listings, will be of critical importance. Exclusive arrangements tying particular EPGs to particular service bundles may become a problem requiring regulatory intervention to ensure third-party access on fair, transparent and non-discriminatory terms" (EC, 1997, p. 24).

U.S. policymakers have grown increasingly concerned about issues of first-screen and presentation bias, although regulatory action has so far been limited.[33] However, with the merger of the two major EPG providers (Gemstar and TV Guide) in 1999, the issue has come under scrutiny from the Antitrust Division of the Department of Justice (DOJ). While the DOJ approved the merger without conditions, it is now investigating Gemstar-TV Guide for abuse of its control over critical patents for EPG services.[34] In Europe, by contrast, regulators have taken a more active role in regulating EPG services, either to protect third-party programmers and service providers or to favor publicly-funded broadcasters. In the UK, for example, OFTEL has interpreted EPGs as covered by the non-discriminatory rules for telecommunications access services,[35] while the ITC has adopted a "code of conduct" for EPG providers that, among other things, mandates that the visual interface grants public service channels "due prominence."[36] As discussed below in the BiB case, European competition authorities have also acted against exclusivity arrangements between EPG providers and dominant platform operators.

## 4. The cases: The AOL/TW merger and British Interactive Broadcasting

The debate about the proper tools and scope of regulatory action vis-à-vis interactive TV services has already surfaced in a number of cases. In this section we analyze two of the most prominent ones: the AOL/Time Warner merger and the British Interactive

---

[33] For example, a few provisions were adopted in the Telecommunications Act of 1996 in the case of EPG services offered by Open Video Systems operators (Section 653(b)), as well as in the Satellite Home Viewer Improvement Act of 1999 (Section 338) for EPGs offered by DBS operators. These provisions however do not extend to cable systems.
[34] "A guide to navigate the TV maze gives pause", *New York Times*, June 25, 2001, p. C1.
[35] See OFTEL (1998). Digital television and interactive services: Ensuring access on fair, reasonable, and nondiscriminatory terms. Available at www.oftel.gov.uk.
[36] Independent Television Commission (1997). *ITC Code of Conduct on Electronic Programme Guides*. London: ITC.

Broadcasting (BiB) case, a joint venture for the launch of interactive TV services in the UK created by BSkyB, British Telecommunications (BT), Midland Bank (part of the HSBC banking group), and Matsushita, the Japanese consumer electronics giant. We contrast the approach taken by U.S. and European regulators to the issues raised by these cases and analyze the implications of the regulatory obligations imposed in each case.

a.  The AOL/Time Warner merger

The January 2000 announcement of the merger between AOL and Time Warner prompted close scrutiny by federal regulators. The FTC concluded that the combination of AOL's Internet properties with Time Warner's cable holdings and content assets had anti-competitive effects in three distinct markets: broadband Internet access service, broadband Internet transport service, and interactive TV services.[37] While most of the policy debate that followed the merger announcement focused on the first two issues, the FTC findings also brought attention to the architecture of next-generation broadcasting networks. Much of the investigation concerning interactive TV centered on AOLTV, AOL's interactive TV product. The existing generation of AOLTV consists of a stand-alone set-top box that connects to a cable or DBS receiver and blends this video programming with interactive content transmitted via a narrowband dial-up modem. While regulators raised few concerns about this service, AOL's plan to upgrade it by embedding AOLTV within cable boxes and utilizing the broadband Internet platform of the cable operator did trouble competition authorities. As the FTC explains,

> "AOL recently launched AOL TV, a first generation ITV service, and is well positioned to become the leading ITV provider. Local cable companies will play the key role in enabling the delivery of ITV services. After the merger, AOL Time Warner will have incentives to prevent or deter rival ITV providers from competing with AOL's ITV service. Thus the merger could enable AOL to exercise unilateral market power for ITV services in Time Warner cable areas, which also affects the ability of ITV providers to compete nationally"[38]

---

[37] Federal Trade Commission (2000). Complaint In the matter of America Online Inc. and Time Warner Inc.. Docket No. C-3989.

[38] Federal Trade Commission (2000). Analysis of proposed consent order in the matter of America Online Inc. and Time Warner Inc.. Docket No. C-3989, p. 2.

Despite the strong wording of these findings, the FTC ultimately imposed rather weak remedies concerning interactive TV. The consent decree simply prohibits AOL-Time Warner from interfering with its subscribers' ability to use the interactive signals or "triggers" provided by programmers that it has agreed (or is forced by statute) to carry.[39] In essence, the FTC order only addressed one of the possible anti-competitive strategies discussed above, that of network operators "stripping" the signals of unaffiliated programmers from its interactive content. Other discriminatory practices related to downstream transmission, upstream transmission (the return path) and the home terminal were left unaddressed.

The FCC concurred with the findings of the FTC merger review vis-à-vis interactive TV:

> "AOL Time Warner would have the potential ability to use its combined control of cable system facilities, video programming and the AOLTV service to discriminate against unaffiliated video programming networks in the provision of ITV services in the provision of ITV services. We also find that AOL Time Warner may have incentives to engage in such discriminatory behavior."[40]

The FCC analysis is broader in scope and acknowledges that the anti-competitive strategies available to AOL/TW go beyond the "stripping" of interactive content of unaffiliated programmers. It also notes that the Memorandum of Understanding by which the merger parties committed to provide customers with a choice of ISPs does not obligate AOL/Time Warner to provide access to the cable broadband platform for interactive TV uses. Nonetheless, the FCC declined to impose additional conditions on the parties pending further examination of market developments and the potential incentives for discriminatory behavior by AOL/TW. In the Commission's analysis, the FTC's prohibition on "stripping" coupled with the conditions relating to the availability of multiple ISPs suffice to protect competition, at least during the initial stages of the interactive TV market. In the words of then Commissioner (and now Chairman) Michael Powell, "although it is possible to hypothesize public interest harms flowing from a cable operator's control of assets like those at issue in this merger, the market is too immature to conclude with any confidence whether such harms are sufficiently probable to warrant direct government intervention."[41]

---

[39] Federal Trade Commission (2000). Decision and Order In the matter of America Online Inc. and Time Warner Inc.. Docket No. C-3989, p. 11.
[40] Federal Communications Commission (2000). In the Matter of Applications for Consent to the Transfer of Control of Licenses and Section 214 Authorizations by Time Warner Inc. and America Online, Inc., Transferors, to AOL Time Warner Inc., Transferee. Memorandum Opinion and Order. CS Docket No. 00-30, p. 90.

b. The British Interactive Broadcasting case

BiB operates one of the largest and most advanced interactive TV service worldwide. It is available to the nearly five million subscribers of BSkyB's digital TV satellite service, offering a variety of dedicated services such as e-mail, electronic banking, games, and gambling, as well as program-related services tied to channels offered by BSkyB. The company was created in 1997 as a joint venture between BSkyB, British Telecommunications (BT), Midland Bank, and Matsushita.[42] It provides interactive TV services in the UK by means of satellite broadcasting (leased from BSkyB, with BT responsible for the uplink) in combination with a narrowband return path through a standard telephone line. The terminal equipment needed to use BiB services is embedded in the BSkyB digital TV set-top box, which BiB partly subsidizes (this includes a proprietary API developed by OpenTV[43] and BSkyB's EPG). Revenues come from end-users and from retailers and interactive TV service providers which BiB carries on its platform.[44]

European competition authorities raised two main concerns about BiB. First, that the company would use its control of the set-top box software components to foreclose competition in interactive TV services, denying third parties access to the heavily subsidized boxes being deployed. Second, that BiB would enhance the already dominant position of BSkyB and BT in the markets for pay-TV and telecommunications local loop respectively. In October 1998, the EC approved the joint venture subject to a number of conditions. In contrast to the AOL/TW case, the regulatory concerns were centered not on the transmission or the return path layers, but on ensuring that "third parties, whether operators of digital television or digital interactive TV services, have fair, reasonable, and non-discriminatory access to all proprietary components of the digital set-top box which BiB will subsidise."[45] This different focus is due to the fact that while cable operators effectively control the transmission infrastructure, satellite TV operators lease capacity from (oftentimes unaffiliated) satellite carriers.[46] Market power therefore stems not from control over transmission infrastructure but rather from first-mover advantages and switching costs associated with proprietary home terminals.[47]

---

[41] Press statement of Commissioner Michael Powell on the approval of AOL/Time Warner merger.
[42] In March 2001, BSkyB completed the acquisition of HSBC's and Matsushita's shareholdings in BiB. It now controls 80.1% of the company, with BT remaining a minority partner.
[43] OpenTV is a developer of interactive TV software originally founded by Sun Microsystems and Thomson Multimedia in 1994.
[44] According to BiB, there are currently 35 retail partners on the platform.

One of the conditions imposed concerned the recovery of the home terminal subsidy. The Commission forced BiB to establish a separate company to manage the subsidy payments in order to ensure that the recovery is evenly distributed among service operators and broadcasters, whether affiliated with BiB and its partners or not. It also demanded that the subsidy was not linked to a subscription to BSkyB's pay-TV service. Another condition related to the terms of access to the home terminal components. BiB agreed to provide, upon request, the API specifications and other proprietary technical systems to third parties. The Commission also forced BiB to end its exclusivity agreement with BSkyB whereby BiB would be the only available interactive TV service on BSkyB's EPG. In addition, the Commission also imposed several obligations on the joint venture partners. BSkyB agreed to offer access services to programmers and interactive TV service providers (including BiB) on fair, reasonable, and non-discriminatory terms regulated by OFTEL under the access control class license.[48] It also agreed to supply, upon request, a "clean feed" (i.e., stripped of interactive applications) of its film and sports channels to other MPVPDs (e.g., cable operators) in order to prevent bundling strategies that would favor BiB. Finally, BT agreed to divest its existing cable interests.[49]

---

[45] McCallum (1999), p. 13.
[46] In the case of BSkyB, it leases satellite capacity from SES (Société Européenne des Satellites).
[47] For a discussion of these advantages see Cave (1997).
[48] Class licence for the running of telecommunication systems for the provision of access control services granted by the secretary of state under section 7 of the telecommunications act 1984. Available at www.dti.gov.uk.

The conditions imposed by the EC on the BiB venture are consistent with the established doctrine among Community competition authorities that ex-post competition rules are insufficient to remedy the problem of access to telecommunications facilities, and thus need to be supplemented by ex-ante, sector-specific regulations.[50] This doctrine has been implemented through a series of Community Directives under the so-called Open Network Provision (ONP) framework, which imposes on telecommunications operators having significant market power certain non-discriminatory obligations that go beyond those that would normally apply under general competition law.[51] It is interesting to note that a few weeks before the BiB decision, the Commission adopted the Access Notice, which explicitly states that the ONP framework extends not only to telecommunications facilities, but to "access issues in digital communications sectors generally."[52]

This extension of telecom-type regulations to next-generation broadcasting technologies as exemplified in the BiB decision has drawn criticism about regulatory overreach. Critics argue that Community competition authorities have taken a narrow, static view of what is still a nascent market, thus discouraging investments and rewarding less innovative firms (e.g., Veljanovski, 1999; Larouche, 1998). The debate draws attention to the fact that the application of competition law necessarily depends on underlying assumptions about how the market should work and what goals should be prioritized. In our view, the assumption that access to network components by third parties is a precondition for long-term innovation is hardly static. Furthermore, it is likely that the nature of broadcasting services demands a more narrow, possibly nation-based definition than in the case of telecommunications (Temple Lang, 1997). Lastly, market developments have simply proved these arguments unfounded. With Britain leading the world in the deployment of interactive TV,[53] it is clear that the condition imposed have hardly discouraged investment in this maturing market.

---

[49] For details on these conditions see Notice published pursuant to Article 19(3) of Regulation 17, OJ C 322 of October 21, 1998.
[50] See Ungerer (2000).
[51] The general framework is provided by the ONP Framework Directive (Council Directive 90/387/EEC, OJ L192/1). The ONP framework has been applied to several industry segments, among them leased lines (92/44/EEC), packet-switched data (EC, 92/382/EEC), and voice telephony (95/62/EC).
[52] Notice on the application of the competition rules to access agreements in the telecommunications sector, OJ C 265, p.3.

## 5. Conclusion

In the aftermath of the AOL/TW merger the debate about open cable access seems to have faded. However, the more general problem of nondiscriminatory access to the basic layers of communications infrastructure (whether cable lines, the local loop, or the digital TV user terminal) is arguably the crucial issue for industry and regulators in the post-convergence era. In this paper we have examined how this problem has re-emerged with the migration to the third generation of broadcasting technology, that of interactive TV. We argued that absent regulatory safeguards that provide for non-discriminatory access to several network components (including digital set-top box components), dominant network operators are likely to leverage ownership of delivery infrastructure into market power over interactive TV services, foreclosing competition and discouraging third parties and users from experimenting with unimagined ways to use television.

In the case of interactive TV, the question of open access is not about extending existing regulatory principles to the new generation of technologies (as it is for the case of broadband Internet). Rather, it is about seizing the opportunities offered by these new technologies to better serve our policy goals. Broadcasting regulation has traditionally taken distribution scarcities and closed network architecture as a fact of life dictated by the available technology, thus relying on ownership rules, licensing criteria, must-carry rules and other regulatory instruments to attain its goals.[54] It is now widely acknowledged that this approach has not only largely failed on its own merits but that it is inadequate for the post-convergence era.[55] The third generation of broadcasting calls for shifting the focus of regulatory action from government "tinkering with the configuration of a mass media market"[56] to rules that ensure nondiscriminatory access to the capacity to experiment with and provide information, entertainment, and transaction services over broadcasting networks.

---

[53] See "Those interactive Britons are tuning on their tellies," The New York Times, April 18, 2001.
[54] See among many others Pool (1983) and Mulgan (1991).
[55] For a U.S. critique see Hazlett and Spitzer (2000). For a European critique see Hoffmann-Riem (1996).

American regulators have so far imposed rather toothless safeguards to prevent discriminatory behavior by incumbent network operators in the interactive TV market. Furthermore, these rules are dispersed across statutes addressing different platforms, thus adversely affecting market competition. European authorities, by contrast, are in the process of fashioning a comprehensive framework that addresses problems of access and interconnection across electronic communications networks.[57] This framework does not impose specific remedies but rather lays out general principles to tackle problems as they arise. By addressing access in a piecemeal, ad-hoc fashion, U.S. policymakers threaten to undermine the very basis of the unprecedented innovation in telecommunications and information services of the last decade and forego the possibility to overhaul an antiquated broadcasting regulatory regime.

---

[56] Benkler (2000), p. 562.
[57] See Amended proposal for a Directive of the European Parliament and of the Council on access to, and interconnection of, electronic communications networks and associated facilities. COM(2001) 369 final.